\documentclass[preprint,12pt]{elsarticle}

\usepackage{amssymb}
\usepackage{amsmath}

\journal{Communications in Nonlinear Science and Numerical Simulation}

\begin{document}

\begin{frontmatter}



\title{Universal Boolean Computation in Negatively
Coupled Circle Map Lattices}


\author[label1,label2]{Priyanka D. Bhoyar, Prashant M. Gade} 

\affiliation[label1]{organization={Department of Physics, Seth Kesarimal Porwal College, Kamptee},
            addressline={Rashtrasant Tukadoji Maharaj Nagpur University}, 
            city={Nagpur},
            postcode={441001}, 
            state={Maharashtra},
            country={India}}
            
\affiliation[label2]{organization={Department of Physics, Ramniranjan   Jhunjhunwala College of Arts,
Science and Commerce},
            addressline={Ghatkopar West},
             city={Mumbai},
             postcode={400086},
             state={Maharashtra},
             country={India}}

\begin{abstract}
We investigate the computational capabilities of a one-dimensional lattice of negatively coupled circle maps in its traveling wave regime. We interpret propagating waves as information carriers and exploit their interactions to implement Boolean logic. The logical states are encoded by the propagation direction of the traveling waves, while the presence or absence of a wave at a designated output location represents the logical output. Using controlled wave collisions, we demonstrate the realization of NAND and NOR gates along with other basic gates. Since NAND and NOR operations are functionally complete, the results establish that the traveling wave regime of the negatively coupled circle-map lattice supports universal Boolean computation in principle. The demonstrated logical operations remain valid over a finite range of the coupling strength $\epsilon$ and map parameter $\omega$, indicating robustness of the computational scheme to parameter variations. Our results therefore demonstrate that computational functionality can emerge from the collective nonlinear dynamics of a coupled map lattice, providing a connection between the dynamical properties of traveling wave states and collision based information processing.
\end{abstract}

\begin{graphicalabstract}
\begin{figure}[h!]
\centering
\includegraphics[width=0.85\linewidth]{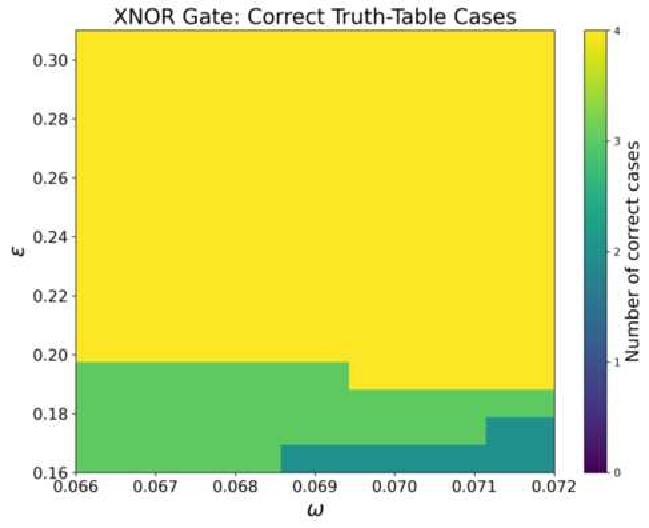}
\end{figure}
\end{graphicalabstract}

\begin{highlights}
\item Demonstrates collision-based logic using traveling waves in a negatively coupled circle-map lattice.

\item Encodes logical inputs through the propagation direction of traveling waves.

\item Exploits wave annihilation and survival to realize Boolean logic operations.

\item Implements NOT, XNOR, AND, OR, XOR, NAND, and NOR logic gates through numerical simulations.

\item Investigates the robustness of logical operations against variations in system parameters.
\end{highlights}


\end{frontmatter}


\section{Introduction}\label{sec1}
Conventional computation is mainly associated with electronic circuits and digital devices. However, physical systems can also process information through their natural dynamics. This idea has motivated the study of unconventional or physical computation, where computation is performed by physical processes rather than conventional electronic components. In particular, collision based computing uses localized structures, particles, pulses, or waves as information carriers, and their interactions implement the logical operations. This idea has been explored in several physical settings, including cellular automata, reaction-diffusion systems, chemical media, optical systems, and mechanical or wave-based systems. In such approaches, information is encoded by the presence, absence, or state of propagating structures, while collisions between them perform various computational operations. For example, in reaction-diffusion computers, the computation is realized by spreading and interaction of diffusive waves. Several computationally universal reaction-diffusion chemical devices such as diodes\cite{dupont1998excitable}, logic gates\cite{toth1995logic,sielewiesiuk2001logical} in Belousov-Zhabotinsky medium and an XOR gate in the Palladium processor\cite{adamatzky2002experimental} have been implemented.
In optical nonlinear media, where localized optical solitons can serve as information carriers and their interactions can produce state transformations. Early work explored computation through collisions of optical solitons and vector solitons, establishing soliton interactions as a possible basis for information processing \cite{jakubowsky2002computing}. Optical logic operations have subsequently been proposed using interactions between localized optical structures, including soliton-dragging mechanisms \cite{McLeod1995}. More recently, energy sharing collisions of bright solitons in the Manakov system have been utilized to construct logical operations such as OR and NOR \cite{Vijayajayanthi2018}, with extended schemes demonstrating NAND, NOR, AND, OR, XOR and XNOR operations \cite{Vijayajayanthi2023}.
Collision based computing has also been explored in mechanical and wave based systems. For example, solitons propagating through one-dimensional mechanical lattices have been shown to interact at lattice junctions in ways that implement Boolean operations such as AND, OR, and NOT \cite{Siccardi2016}. More generally, collision based computation can be realized through interactions of mobile localized structures whose trajectories,  states, or post collision dynamics encode logical information \cite{Adamatzky2002}.

This concept has been extensively explored in cellular automata. Rennard constructed AND, OR, and NOT operations in Conway’s Game of Life through appropriate interactions of propagating gliders and demonstrated their use in a binary adder \cite{Rennard2002}.
Zhang and Adamatzky demonstrated that collision based computation can be implemented in an excitable cellular automaton, using localized propagating structures as information carriers. In their approach, logical information is encoded by the presence or absence of excitation, while collisions between propagating excitation patterns perform logical operations. They showed that appropriate arrangements of these collisions can realize the components of a two-bit adder \cite{Zhang2009}.
Evolutionary approaches have further been used to identify cellular automata capable of supporting collision based logic \cite{Sapin2008}. Such an approach motivates the investigation of collision based computation in other spatially extended dynamical systems, including coupled map lattices.

A particularly interesting example is provided by the one-dimensional lattice of coupled circle maps studied by Gade et al.\cite{Gade2007}. They considered a lattice of circle maps with repulsive (negative) coupling and showed that the system supports traveling wave excitations and investigated their persistence properties. The existence of such propagating structures in a relatively simple discrete dynamical system makes this model an appealing candidate for exploring wave based information processing. While the work of Gade et al. was concerned with the dynamical properties and persistence of traveling waves, the present work asks a different question: can these naturally occurring traveling wave excitations be used as carriers of logical information?

In the present work, we use the traveling wave dynamics of the negatively coupled circle map lattice as a basis for collision based logic. Logical inputs are encoded by the propagation direction of traveling waves. A right-moving wave (R) represents logical $0$, whereas a left-moving wave (L) represents logical $1$. Logical operations are realized through interactions between these waves, with collisions potentially leading to wave annihilation. The logical state is determined by the presence or absence of a wave at a designated output site.
We demonstrate NOT, XNOR, XOR, OR, AND, NOR and NAND logic gates through numerical simulations.
The realization of NAND and NOR is particularly significant because each of these gates is computationally universal. Thus, the results go beyond demonstrating isolated logical operations and indicate that the same underlying nonlinear lattice dynamics can support universal Boolean computation.

\section{Model}
We investigate logical computation using a one-dimensional lattice of negatively coupled circle maps, based on the model introduced by Gade et al. \cite{Gade2007}
We consider a one-dimensional lattice of size $L=90$ or $150$ sites, where each site $i$ is associated with a phase variable $x_i(t)\in[0,1)$. The dynamics is governed by a coupled circle map with repulsive (negative) nearest-neighbor coupling,

\begin{equation}
x_i(t+1)= f[x_i(t)]-\frac{\epsilon}{2} [x_{i-1}(t)+x_{i+1}(t)-2x_i(t)]
\end{equation}

where, $f(x)= x+\omega-\frac{k}{2\pi}\sin(2\pi x)$ is the standard circle map, and $\epsilon>0$ denotes the strength of the repulsive coupling. Periodic boundary conditions are imposed on the lattice as $x_0=x_L$, $x_{L+1}=x_1$. The dynamics is confined to the interval [0,1] using the following rule. If $\it{int}[x_i(t)]=m$, then
\begin{equation*}
x_i(t)=
\begin{cases}
x_i(t)-m, & x_i(t)\geq 0,\\[4pt]
x_i(t)-m+1, & x_i(t)<0.
\end{cases}
\end{equation*}
For suitable values of the repulsive coupling strength $\epsilon$, the coupled circle map exhibits a traveling wave regime in which localized phase excitations propagate through the lattice. For $k=1$ and $\omega=0.068$, Gade \textit{et al.} observed traveling waves for an intermediate range of coupling, approximately $0.155\le\epsilon\le0.36$. The negative coupling favors phase differences between neighboring sites, allowing disturbances to propagate as left-moving and right-moving waves. These waves can interact when oppositely directed excitations meet, with the interaction resulting in either transmission or annihilation depending on the dynamical state of the interacting waves. This propagating wave dynamics provides the basis for using wave interactions as information carriers in logical operations.

\section{Results and Discussion}
The input is encoded by the direction of the travelling wave, with a right-moving wave (R) representing logical $0$ and a left-moving wave (L) representing logical $1$.
To generate localized traveling wave excitations in the circle map lattice, a localized three-site perturbation is then introduced at the desired lattice position. Two orientations of the perturbation are used to generate waves propagating in opposite directions. Specifically, for a wave centered at lattice site $i$, the initial states are chosen as $(x_{i-1},x_{i},x_{i+1})$ =$(0.95,,0.26,,0.10)$, for a left-moving wave, whereas
$ (x_{i-1},x_{i},x_{i+1})$ = $(0.10,$ $0.26,0.95)$ generates a right-moving wave. All other lattice sites are initialized at the fixed-point value $x^*=\frac{1}{2\pi} \sin^{-1}\left(\frac{2\pi\omega}{k}\right)$. The subsequent evolution of these localized perturbations under the negatively coupled circle map dynamics produces traveling wave excitations.
The circle map parameters are fixed at $\omega=0.068$ and $k=1.0$, while the coupling parameter is chosen as $\epsilon=0.25$.
We construct Boolean logic operations including NOT, XNOR, AND, OR, NOR, and NAND, demonstrating how computational functionality can emerge directly from the collective dynamics of the coupled map lattice.

\subsection{NOT operation}
\begin{figure}[h!]
\centering
\includegraphics[width=\linewidth]{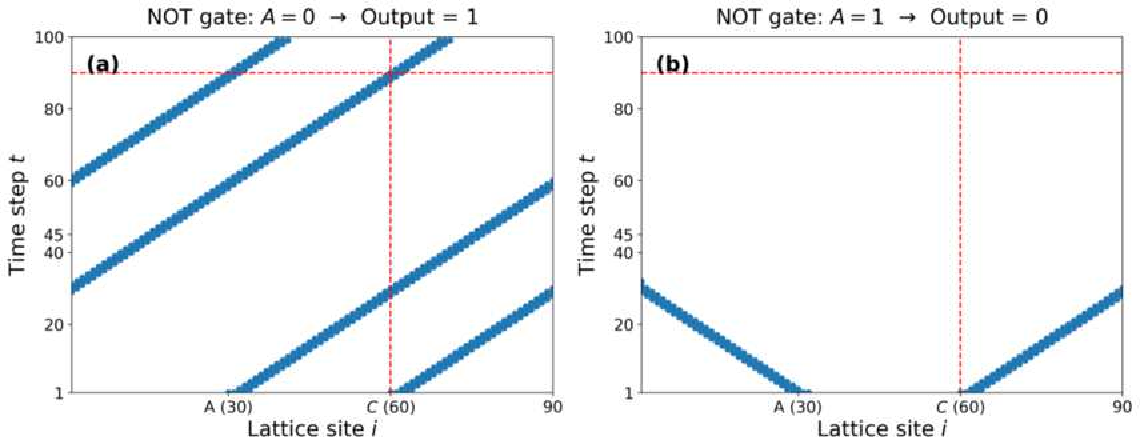}
\hfill
\caption{ Shows the spatiotemporal evolution of the traveling waves for the NOT gate. (a) For input $A=0$, the reference wave propagates without annihilation and represents output $1$ at $i=60,t=90$. (b) When $A=1$, the input and reference waves collide and annihilate, resulting in output state $0$.}
\label{fig:not}
\end{figure}

We simulate a lattice of size $L=90$ for $t=100$ time steps under periodic boundary conditions. A right-moving or left-moving traveling input wave $A$ is introduced at site $30$ and a right-moving reference wave $C$ is introduced at site $60$. The output is determined from the survival of the reference wave. For a right-moving traveling wave input, the reference wave continues to propagate and reaches the output location  $i=60$, at $t=90$ (see Fig.\ref{fig:not}(a)).

\begin{table}[h!]
\centering
\begin{tabular}{|c|c|c|c|}
\hline
\textbf{Input }A & \textbf{Input wave at A} & \textbf{Output detector at }C & \textbf{Output} \\
\hline
0 & \text{R} & \text{Reference wave survives} & 1 \\
1 & \text{L} & \text{Reference wave annihilated} & 0 \\
\hline
\end{tabular}
\caption{Truth table for NOT operation. The input is encoded by the direction of the traveling wave: Left-moving (L) and right-moving (R). The output is determined by the survival of the reference  wave at the control site $C$.}
\label{tab:not}
\end{table}

For the left-moving traveling wave input $A=1$, the input wave collides with the reference wave and annihilates it as shown in Fig.\ref{fig:not}(b). Therefore, no reference wave reaches the output detector. The output is therefore determined by the survival of the control wave as shown in the truth table \ref{tab:not}.

\subsection{XNOR operation}
\begin{figure}[h!]
\centering
\includegraphics[width=\linewidth]{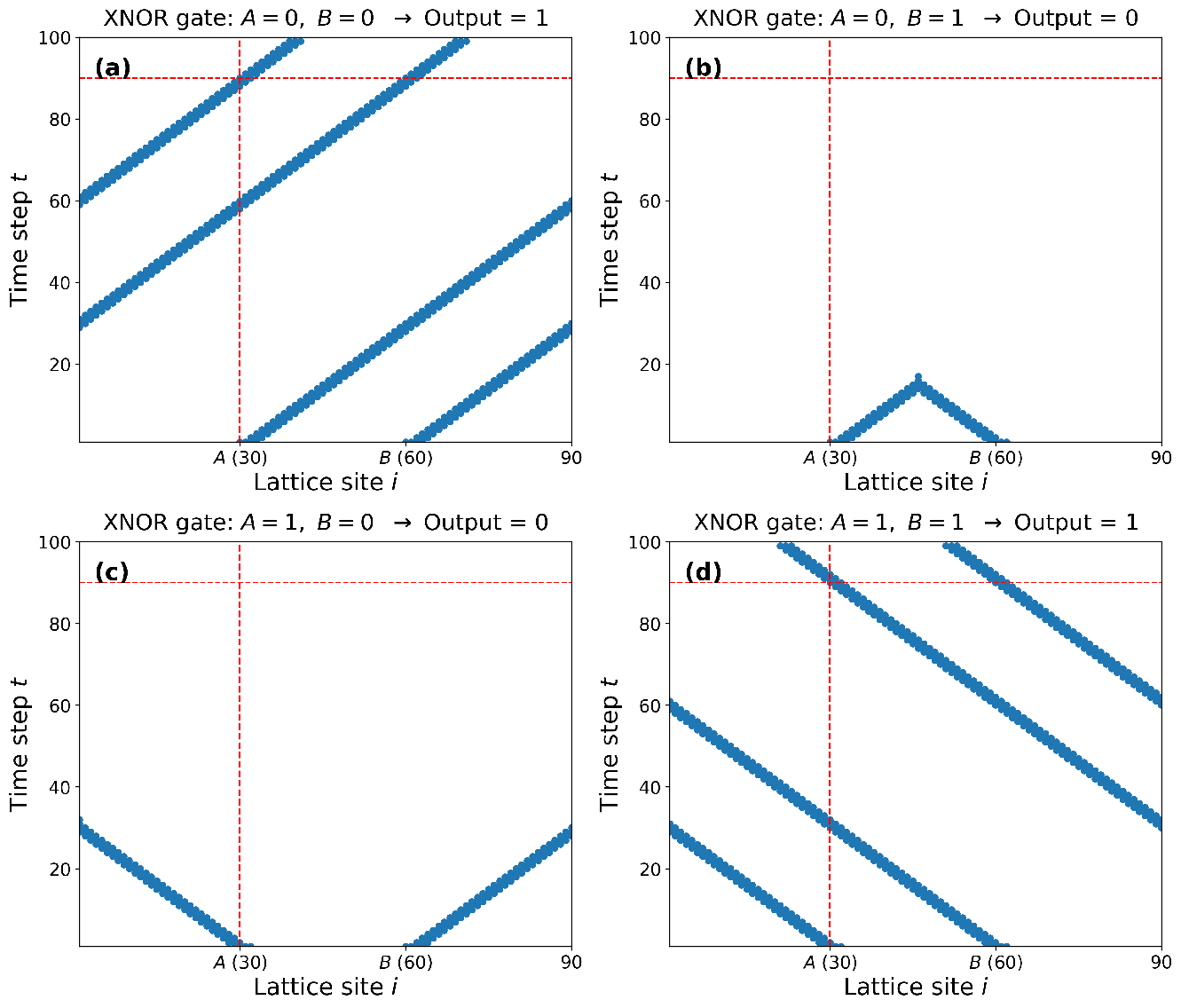}
\hfill
\caption{Shows the spatiotemporal evolution of the traveling input waves for the XNOR gate.
In case (a) $A=B=0$ and (d) $A=B=1$, the input waves propagate without annihilation and produce output $1$ at the output detector $A$ at $t=90$. In case (b) $A=0, B=1$ and $A=1, B=0$, both the input waves are counter propagating, hence collide and annihilate, resulting in output $0$. }
\label{fig:xnor}
\end{figure}

The Exclusive NOR or XNOR operation emerges naturally from the intrinsic interaction rules of the traveling waves in the negatively coupled circle map lattice. The propagation and annihilation rules of the model directly reproduce the XNOR truth table without requiring an additional logical transformation. We simulate a lattice of size $L=90$ with periodic boundary conditions for 100 time steps. The two input traveling waves $A$ and $B$ are introduced at sites $30$ and $60$, respectively. When both inputs are in the same logical state, the resulting wave configuration produces an output corresponding to logical $1$. In contrast, when the two inputs are different, their counter propagating waves interact and annihilate, resulting in logical $0$, as shown in Fig.\ref{fig:xnor}. The resulting output for several input combinations is given in truth table \ref{tab:xnor}.

\begin{table}[h!]
\centering
\begin{tabular}{|c|c|c|c|c|c|}
\hline
A & B & \textbf{Input A} & \textbf{Input B} & \textbf {Output detector at A }& \textbf{Output} \\
\hline
0 & 0 & \text{R} & \text{R} & Input wave survives& 1 \\
0 & 1 & \text{R} & \text{L} & Input wave annihilates & 0 \\
1 & 0 & \text{L} & \text{R} & Input wave annihilates & 0 \\
1 & 1 & \text{L} & \text{L} & Input wave survives &1 \\
\hline
\end{tabular}
\caption{Truth table for the XNOR gate. The output can be determined by the survival of either of the input waves. In the present case, the input wave $A$ survives when both the inputs are at same logical state.}
\label{tab:xnor}
\end{table}

Strictly speaking, XNOR (or XOR) and NOT together are not a universal set of logic gates. They can generate only affine Boolean functions and cannot, for example, construct the AND operation. Therefore, while the realization of XNOR/XOR and NOT gates demonstrates useful information processing through traveling wave interactions, an additional nonlinear gate such as AND or OR is required to achieve universal Boolean computation. However, demonstrating NAND and NOR gates establishes functional completeness for Boolean computation within the proposed scheme.

\subsection{OR and AND operation}
A basic OR gate is implemented by using two left-moving reference waves,
$C1$ and $C2$ (located at $ i=60$ and $90$, respectively), placed between the two input waves $A$ and $B$ (located at $i=30$ and $120$ respectively).

\begin{figure}[h!]
\centering
\includegraphics[width=\linewidth]{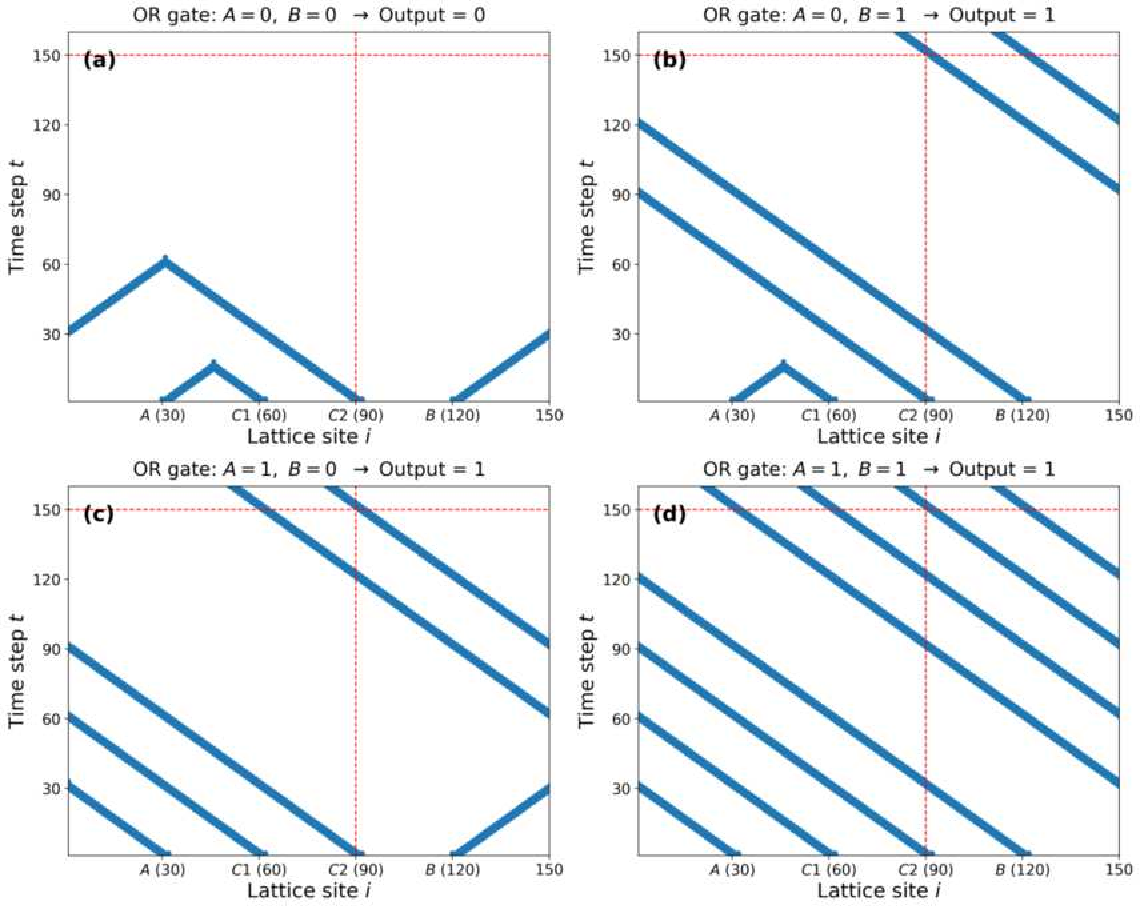}
\hfill
\caption{Shows the spatial realization of the OR gate through traveling wave propagation and collision dynamics. The four panels correspond to the input combinations $(A,B)=(0,0),(0,1),(1,0)$, and $(1,1)$, with the resulting wave configuration at $i=90$ determining the logical output. }
\label{fig:or}
\end{figure}

\begin{table}[h]
\centering
\begin{tabular}{|c| c| c| c| c| c|}
\hline
$A$ & \textbf{Reference} $C1$ & \textbf{Reference} $C2$ & $B$ &
\textbf{Output detector at} $C2$ & \textbf{Output} \\
\hline
0 & L & L & 0 & Reference wave Annihilates & 0\\
0 & L & L & 1 & Reference wave Survives & 1 \\
1 & L & L & 0 & Reference wave Survives & 1 \\
1 & L & L & 1 & Reference wave Survives & 1 \\
\hline
\end{tabular}
\caption{Truth table for the OR gate implemented using traveling wave
collisions. Here, the reference wave $C2$ survives at the output detector site for all cases except when both the inputs are at logical state $0$.}
\label{tab:or}
\end{table}

The output is detected from the survival of the control wave $C2$ at
lattice site $90$ and time step $t=150$. For $A=B=0$,  right-moving input $A$ and $B$
causes the annihilation of both the control waves. Thus, the corresponding output is logical $0$. When either input is left-moving, the corresponding left-moving input wave interacts with the control waves, and the output wave at $C2$ is present at the detection point. Thus,
the output is $1$ whenever at least one of the inputs is $1$ (See Fig.\ref{fig:or}).
The four input combinations and the corresponding output are summarized in Table~\ref{tab:or}.

\begin{figure}[h!]
\centering
\includegraphics[width=\linewidth]{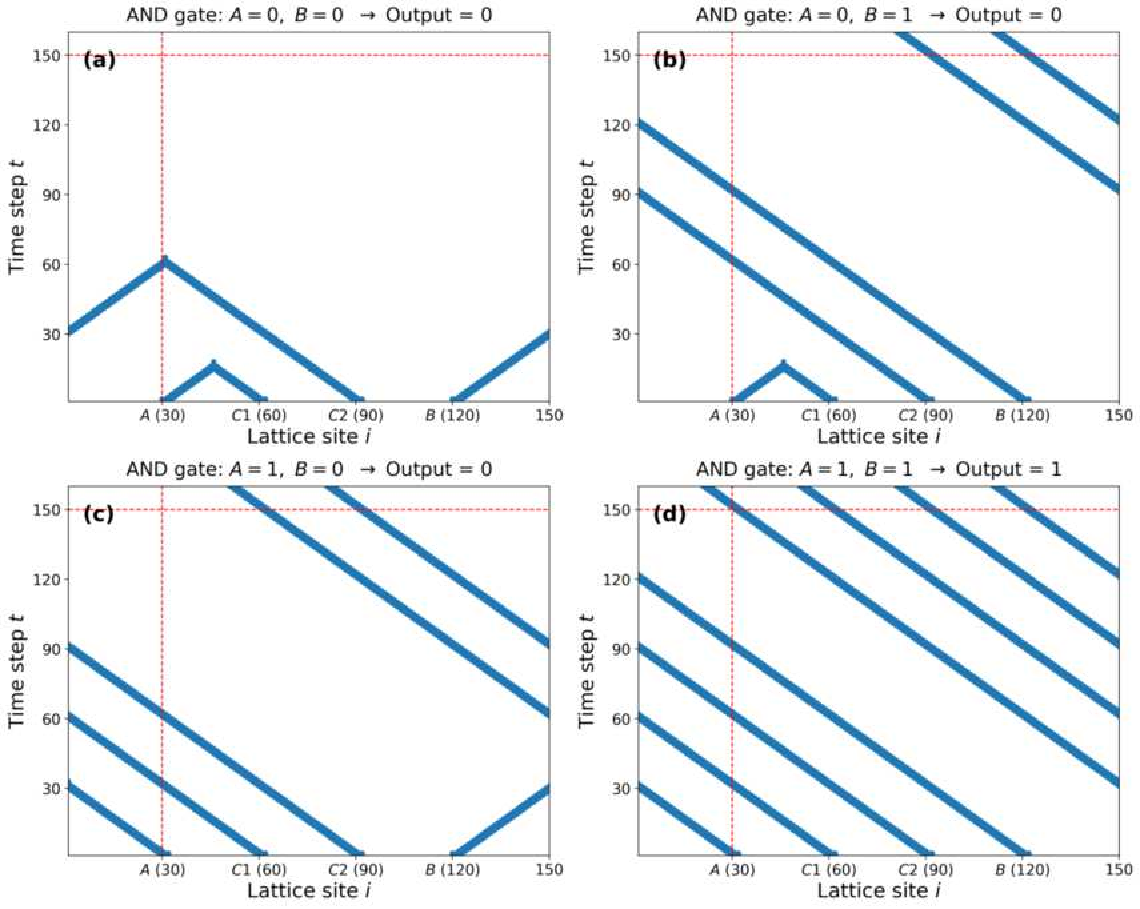}
\hfill
\caption{Spatiotemporal evolution of the traveling waves for the AND gate. The input wave at $i=30$ survives only when there is no interaction between the input and reference waves. In other cases, the $A$ wave annihilates with the counter prapogating wave. }
\label{fig:and}
\end{figure}

\begin{table}[h]
\centering
\begin{tabular}{|c| c| c| c| c| c|}
\hline
$A$ & Reference $C1$ & Reference $C2$ & $B$ &
Output detector at $A$ & Output \\
\hline
0 & $L$ & $L$ & 0 & Input wave annihilate & 0\\
0 & $L$ & $L$ & 1 & Input wave annihilate & 0 \\
1 & $L$ & $L$ & 0 & Input wave annihilate & 0 \\
1 & $L$ & $L$ & 1 & Input wave survives & 1 \\
\hline
\end{tabular}
\caption{Truth table for the AND gate. The output is detected at $A$ at time=150. The input wave survives only when both the inputs are at logical $1$.}
\label{tab:and}
\end{table}

In a similar manner, the AND gate output is realized with the same set of reference and input waves, but an output detector at $A$ at time step=150. In this case, the input wave survives only when the inputs and reference wave are at the same logical level. In all the other cases, the input traveling wave annihilates with the counter propagating waves as shown in Fig. \ref{fig:and}. The truth table is shown in \ref{tab:and}.

\subsection{NAND and NOR operation}
The spatial arrangement of input waves and reference waves is the same as that in the case of OR/AND gates. The NAND/NOR operations are realized by employing two independent right-moving reference waves, $C1$ and $C2$, at the control sites $60$ and $90$0, respectively.

\begin{figure}[h!]
\centering
\includegraphics[width=\linewidth]{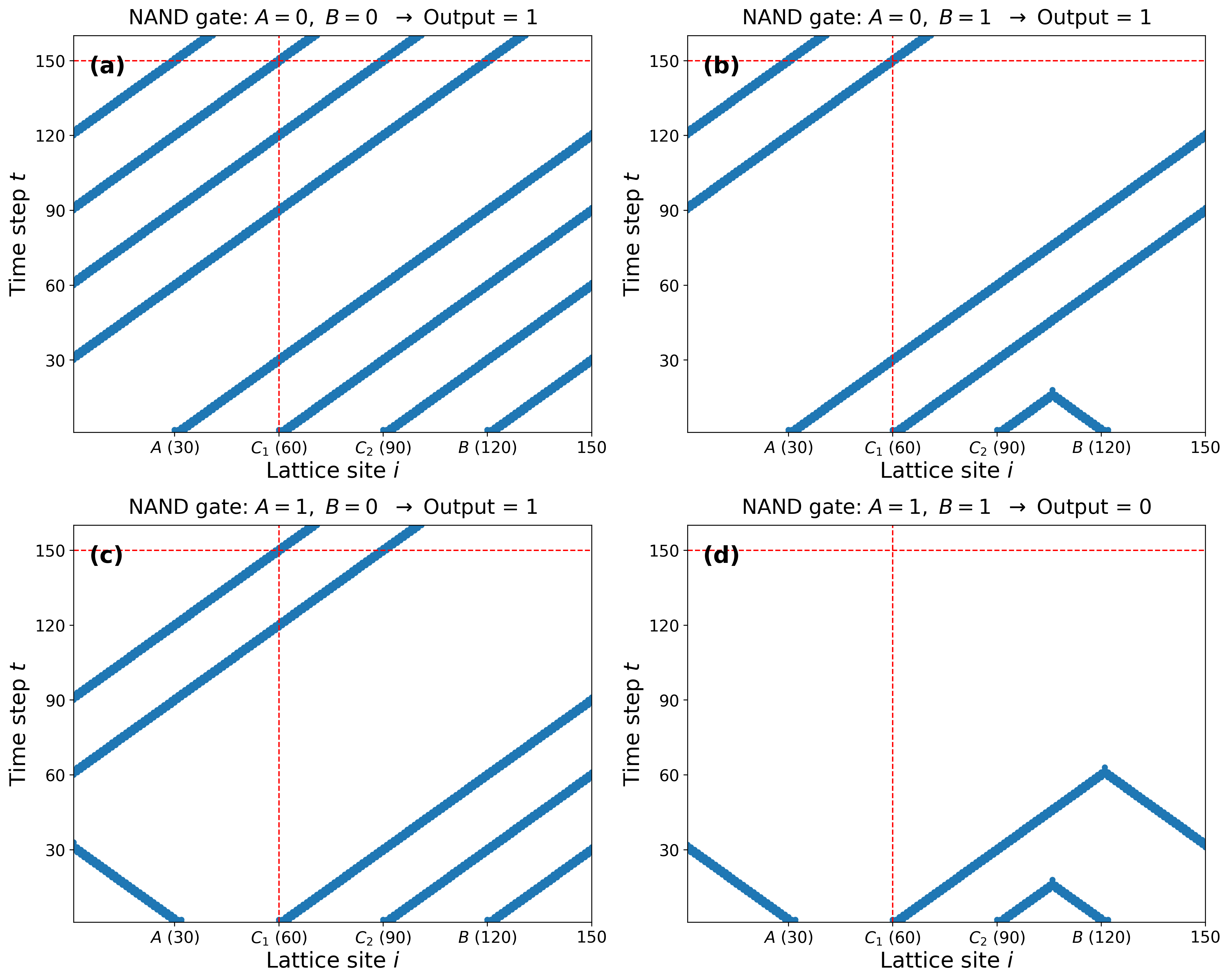}
\hfill
\caption{The four panels correspond to the input combinations $(A,B)$ = $(0,0)$, $(0,1)$, $(1,0)$, and $(1,1)$, demonstrating the corresponding NAND outputs. For $A=1$ and $B=1$, the counter propagating waves interact and annihilate $C1$, resulting in the logical output $0$. In all the other cases, reference wave $C1$ survives.}
\label{fig:nand}
\end{figure}

Fig.\ref{fig:nand} shows the spatiotemporal evolution of the traveling wave inputs $A$, $B$ and reference waves $C1$, $C2$. For $A=1$ and $B=1$, the counter propagating waves interact and annihilate at the collision site, resulting in the logical output $0$. For all other input combinations, the reference wave $C1$ survives, yielding the logical output $1$.

\begin{table}[h]
\centering
\begin{tabular}{|c| c| c| c| c| c|}
\hline
$A$ & \textbf{Reference} $C1$ &\textbf{ Reference} $C2$ & $B$ & \textbf{Output at} $C1$ & \textbf{Output} \\
\hline
0 & R & R & 0 & reference wave survives & 1 \\
0 & R & R & 1 & reference wave survives & 1 \\
1 & R & R & 0 & reference wave survives & 1 \\
1 & R & R & 1 & reference wave annihilated & 0 \\
\hline
\end{tabular}
\caption{Truth table for the NAND operation. Here, the reference waves $C1$ and $C2$ are right-moving. The output is determined from the survival of reference wave $C1$ at time step $=150$. The two left-moving inputs collide with two right-moving reference waves, yielding output $0$. In the other cases, $C1$ survives. }
\label{tab:nand_gate}
\end{table}
Truth table \ref{tab:nand_gate} shows the realization of the NAND operation using traveling wave propagation and collision dynamics.

In the same spatial arrangement of input and reference waves, the NOR operation can be realized by selecting the input wave $B$ as output detector at time step $=150$.

\begin{figure}[h!]
\centering
\includegraphics[width=\linewidth]{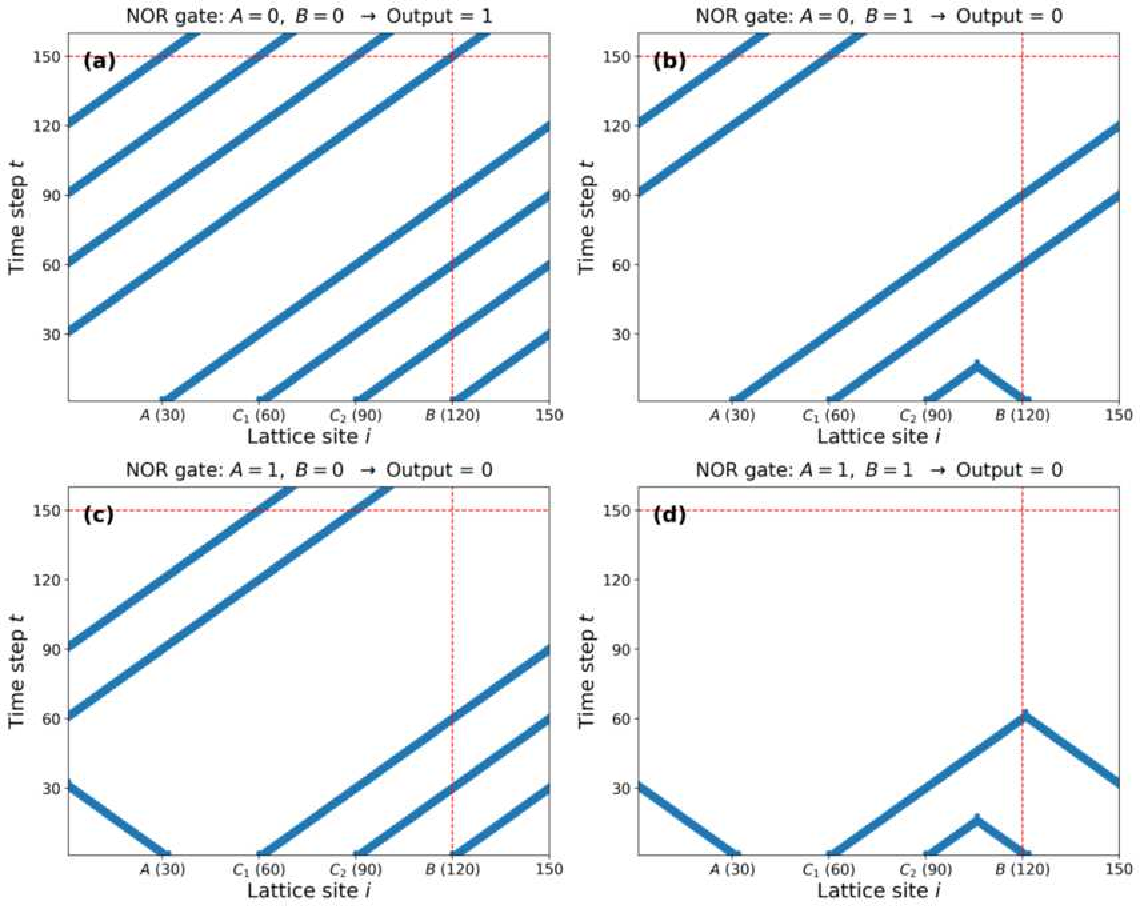}
\hfill
\caption{Spatiotemporal evolution of the traveling waves for the NOR gate. The output is determined by the survival of input wave $B$ at $t=150$. The input wave $B$ survives for $A=B=0$ and annihilates with the counter propagating $C2$ and $A$ in other cases.}
\label{fig:nor}
\end{figure}

In case $A=0$ $B=0$, right-moving $B$ wave survives producing output state $1$. In the other cases, $B$ wave annihilates when it collides with the neighboring wave. The truth table \ref{tab:nor_gate} shows the realization of the NOR operation using traveling wave propagation and collision dynamics

\begin{table}[h]
\centering
\begin{tabular}{|c |c| c| c| c| c|}
\hline
$A$ & Reference $C1$ & Reference $C2$ & $B$ & Output Detector $B$ & Output \\
\hline
0 & R & R & 0 & Input wave survives & 1 \\
0 & R & R & 1 & Input wave annihilated & 0 \\
1 & R & R & 0 & Input wave annihilated & 0 \\
1 & R & R & 1 & Input wave annihilated & 0 \\
\hline
\end{tabular}
\caption{Truth table for the NOR gate implemented using traveling wave collisions. The input wave $B$ survives for $A=B=0$ and annihilates with the counter propagating $C2$ and $A$ in other cases. }
\label{tab:nor_gate}
\end{table}

The realization of both NAND and NOR gates is particularly significant because each of these gates is functionally complete and can, in principle, be used to construct any Boolean logic circuit. Thus, the demonstrated wave-based operations are not limited to the individual gates presented here, but provide a basis for constructing more complex computational elements.

\subsection{XOR operation}
We implement the XOR operation by cascading the XNOR operation with a NOT operation. Since XNOR produces a logical $1$ when the two inputs are identical and a logical $0$ when they are different, applying a NOT operation to its output reverses these states and consequently produces the XOR truth table.
The XNOR output (i=30) propagates to the NOT stage, where it interacts with the reference wave (i=60) using the same collision mechanism described previously. For equal inputs, the XNOR stage produces an output $1$ (fed as left-moving wave to NOT gate at i=30), which is converted to $0$ by the NOT stage. For unequal inputs, the XNOR output is $0$ (fed as right-moving wave to NOT gate at i=30, and the absence of the XNOR wave allows the reference wave to survive, producing an output of $1$. Thus, the cascaded dynamics realizes the XOR operation. This demonstrates that the proposed collision-based gates can be cascaded to construct additional logical operations rather than being restricted to isolated gate operations.

\begin{figure}[h!]
\centering
\includegraphics[width=\linewidth]{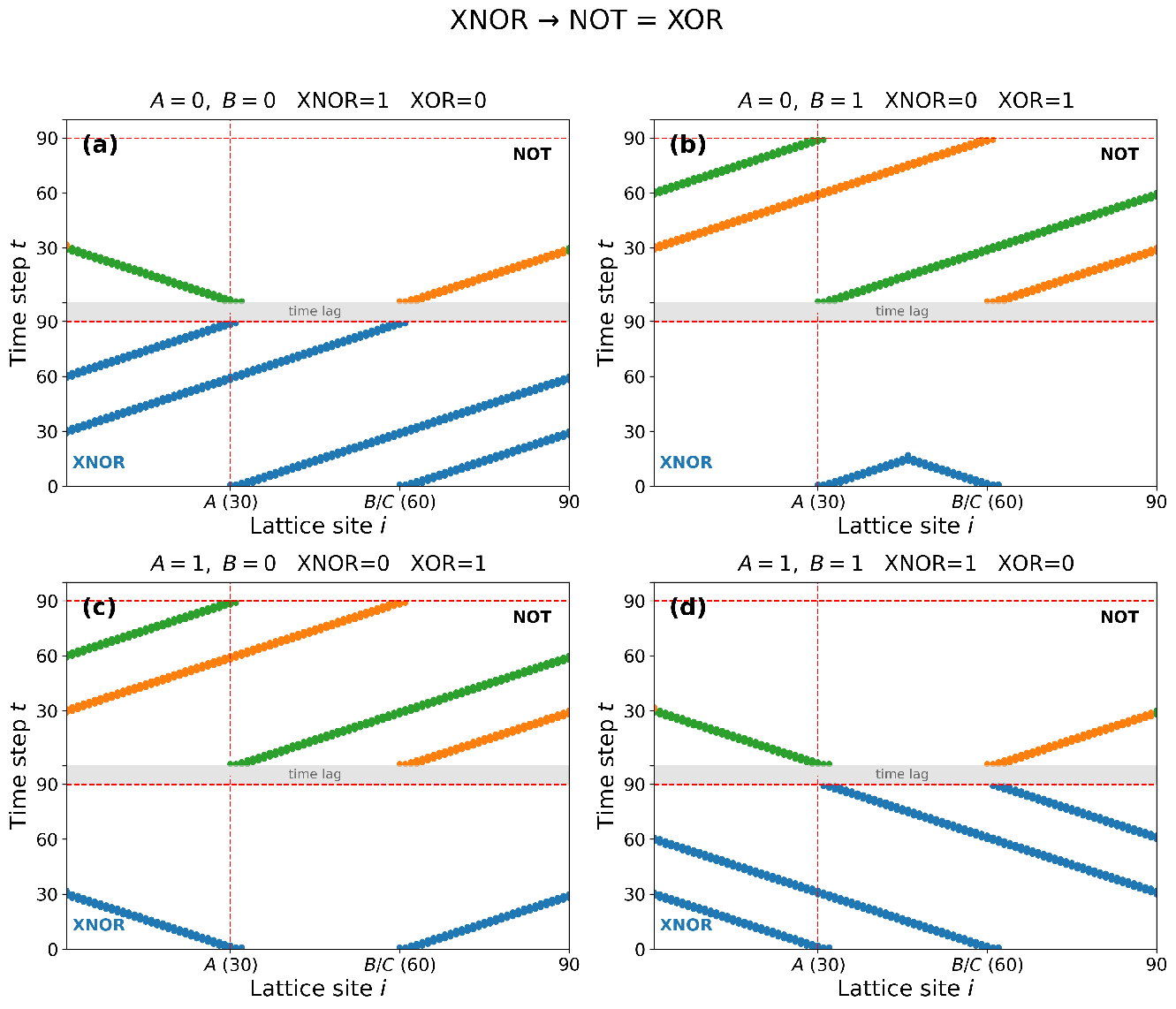}
\hfill
\caption{Spatiotemporal realization of the XOR operation through cascading an XNOR gate with a NOT gate. The four panels correspond to the input combinations (a) (A, B)=(0,0), (b) (0,1), (c) (1,0), and (d) (1,1). The lower part of each panel (($0 \leq t \leq 90$)) shows the spatiotemporal evolution of the XNOR stage, while the upper part (($0 \leq t \leq 90$)) shows the subsequent NOT stage after a finite time delay. The XNOR output is transferred to the NOT stage as the input wave, thereby forming a cascaded dynamical operation. }
\label{fig:xor}
\end{figure}

\begin{table}[h!]
\centering
\begin{tabular}{|c|c|c|c|}
\hline
\textbf{Input A} & \textbf{Input B} & \textbf{Output of} XNOR  & \textbf{XOR}=$\overline{XNOR}$ \\
\hline
0 & 0  & 1 & 0 \\
0 & 1  & 0 & 1 \\
1 & 0  & 0 & 1 \\
1 & 1  & 1 & 0 \\
\hline
\end{tabular}
\caption{Truth table for the XOR gate. The output of the XNOR gate is transferred to a NOT gate as input, yielding the output of the XOR gate.}
\label{tab:xor_gate}
\end{table}

In Fig.\ref{fig:xor}, blue trajectories represent the traveling wave dynamics in the XNOR stage, while the green trajectories denote the propagated XNOR output entering the NOT stage. The orange trajectories represent the reference/control wave used in the NOT operation. The red dashed lines indicate the interface between the two stages and the relevant lattice/time locations, while the shaded region denotes the time lag introduced between the XNOR output and its subsequent processing by the NOT gate. For equal inputs, (0,0) and (1,1), the XNOR stage produces logical 1, which is inverted by the NOT stage to give the XOR output 0. For unequal inputs, (0,1) and (1,0), the XNOR stage produces logical 0, which is subsequently converted to logical 1 by the NOT stage. The results are summarized in truth table  \ref{tab:xor_gate}.

The present study is based on numerical simulations of the coupled circle map lattice. Nevertheless, the underlying mechanism suggests a possible route toward experimental realization. In particular, the discrete nonlinear dynamics and inhibitory coupling considered here could, in principle, be implemented using networks of coupled electronic or other nonlinear dynamical elements. Such an implementation could provide a physical realization of the traveling wave excitations and their collision-based interactions studied here. Experimental investigation of these possibilities may therefore provide an interesting direction for future work.

\subsection{Parameter Robustness}

For a dynamical system to serve as a reliable logical computing element, its logical operation should not depend critically on a single precisely tuned set of dynamical parameters. In collision based computing, this is particularly important because the logical output is determined by the propagation and interaction of traveling waves, which can be sensitive to changes in the underlying dynamical regime. Demonstrating robustness therefore provides evidence that the observed logical operation is a dynamical property of the system rather than a consequence of fine parameter tuning.

\begin{figure}[h!]
\centering
\includegraphics[width=0.65\linewidth]{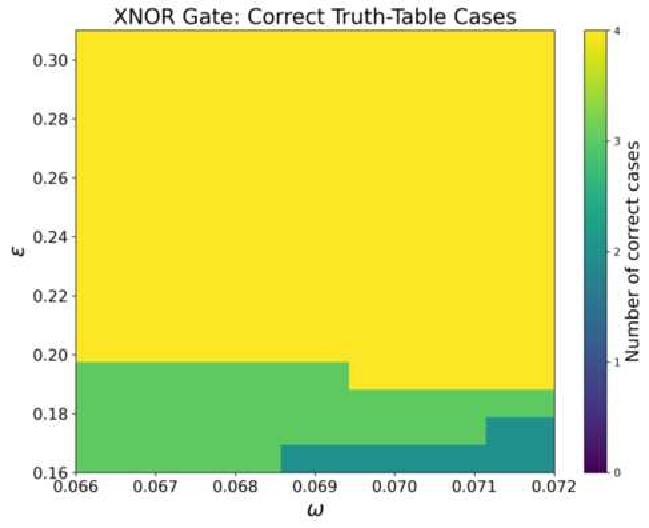}
\hfill
\caption{ Parameter-space map showing the number of correctly reproduced input-output combinations of the XNOR gate as a function of the coupling strength $\epsilon$ and the map parameter $\omega$ over a range where traveling waves are obtained in the lattice. The value indicates the number of correct cases out of the four possible input combinations $(A, B)=(0,0),(0,1),(1,0),(1,1)$. A value of 4 indicates that the complete XNOR truth table is correctly reproduced.}
\label{fig:robust}
\end{figure}

We examine the robustness of the XNOR gate. In this case, the map parameter $\omega$ and the coupling strength $\epsilon$ are varied over a range that yields travelling waves. For each parameter set, all four possible input combinations, (A, B)=(0,0),(0,1),(1,0),(1,1), were simulated and the resulting outputs were compared with the XNOR truth table. The number of correctly reproduced input-output combinations recorded for each parameter set, with a value of four indicating that the complete XNOR truth table was reproduced correctly, is shown in Fig.\ref{fig:robust}.
The gate remains correctly functional over the tested ranges of $\omega=0.066$–-$0.072$ and $\epsilon \sim 0.2$–-$0.35$, when the other parameters are kept fixed. Thus, within the parameter ranges investigated, the logical operation does not require a single finely tuned parameter value. This finite operating region demonstrates the XNOR gate's parameter robustness and supports the reliability of the collision-based logic mechanism in the negatively coupled circle map lattice.

\section{Conclusion}
The ability of a dynamical system to process information is an intriguing example of how collective physical behavior can give rise to functionality beyond the dynamics of its individual elements. In the present work, we explore this possibility in a one-dimensional lattice of negatively coupled circle maps, where traveling waves provide mobile carriers of information and their interactions serve as the basic computational events. The propagation and collision of these excitations enable different logical responses to emerge naturally from the underlying nonlinear dynamics, leading to the realization of NOT, NXOR AND, OR, NOR, NAND and XOR operations. The realization of both NOR and NAND is particularly noteworthy, as either gate by itself is sufficient to generate arbitrary Boolean logic, thereby establishing a route towards universal computation within the dynamical system.
Furthermore, a parameter-robustness analysis of the XNOR gate shows that the logical operation is preserved over a finite range of the map parameter $\omega$ and coupling strength $\epsilon$. This indicates that the demonstrated computation does not rely on a single precisely tuned parameter set and remains stable within a finite operating region of the dynamical system.
More broadly, our results illustrate that computation need not be viewed solely as a process implemented by conventional electronic components, but can also emerge from the spatiotemporal organization of nonlinear physical systems. This perspective opens the possibility of exploring coupled dynamical systems as alternative computational media, where information processing is encoded directly in the evolution and interaction of collective excitations. By appropriately cascading and interconnecting the gates, one can construct arithmetic circuits such as half-adders and full-adders, as well as other combinational logic circuits.

\section*{Acknowledgements}
PMG thanks IMSc, Chennai, for hosting a visit and Prof. Sitabhra Sinha for discussions.
PDB thanks Rashtrasant Tukadoji Maharaj
Nagpur University for providing financial assistance (RTMNU/RDC/2024/242)

\section*{AI disclosure}
During the preparation of this work the authors used [Meta AI / ChatGPT / etc.] for language editing and to improve readability. After using this tool, the authors reviewed and edited the content as needed and take full responsibility for the content of the publication.

\section*{Competing interests}

\section*{Author contributions}
First Author: Conceptualization, research idea, methodology, numerical analysis, interpretation of results, and manuscript preparation.
Second Author: Simulation development, numerical analysis, interpretation of results, manuscript drafting, and manuscript revision.

\section*{Data availiability}
The data supporting the findings of this study are available from the first author upon reasonable request.

\end{document}